\documentclass[useAMS,usenatbib]{mn2e}

\usepackage{graphicx}

\title{Accelerated expansion from cosmological holography}
\author[Maurice, H.P.M. van Putten]
{Maurice H.P.M. van Putten$^1$ \\ 
$^{1}$Astronomy and Space Science, Sejong University, 98 Gunja-Dong Gwangin-gu, Seoul 143-747, Korea}

\begin{document}

\date{}

\pagerange{\pageref{firstpage}--\pageref{lastpage}} \pubyear{2002}

\maketitle

\label{firstpage}

\begin{abstract}
It is shown that holographic cosmology implies an evolving Hubble radius $c^{-1}\dot{R}_H =  -1  + 3\Omega_m$ in the presence of a dimensionless matter density $\Omega_m$ scaled to the closure density $3H^2/8\pi G$, where $c$ denotes the velocity of light and $H$ and $G$ denote the Hubble parameter and Newton's constant. It reveals a dynamical dark energy and a sixfold increase in gravitational attraction to matter on the scale of the Hubble acceleration. It reproduces the transition redshift $z_t\simeq 0.4$ to the present epoch of accelerated expansion and is consistent with $(q_0,(dq/dz)_0)$ of the deceleration parameter $q(z)=q_0+(dq/dz)_0z$ observed in Type Ia supernovae.
\end{abstract} 


\section{Introduction}

General relativity describes a four-covariant geometric theory of gravitation with an exception record in describing
linearized gravity in systems much smaller than the Universe, notably in our solar system (e.g. \cite{bin87}) and 
compact binaries evolving by gravitational radiation losses \citep{hul75}. It hereby passes crucial
observational tests in the embedding of Newton's theory of gravitation by a mixed elliptic-hyperbolic system of equations
(e.g. \cite{van96}), parameterized by Newton's constant $G$ and the velocity of light $c$. The unit of luminosity 
$L_0 = {c^5}/{G}=3.36\times 10^{59}$ herein defines a characteristic scale for the final phase of black hole evaporation 
and, respectively, gravitational radiation from the coalescence of black hole binaries.

In modern cosmology, however, general relativity is faced with a mysterious cosmological constant $\Lambda$
or dark energy and dark matter with no apparent microphysical origin \citep{rie98,per99,rie04}. In describing the large scale 
structure and evolution of the Universe described by the Friedmann-Robertson-Walker (FRW) equations of motion, 
they arise at weak gravity on the scale of the Hubble acceleration $a_H=cH_0$ defined by the cosmological horizon, 
where $c$ denotes the velocity of light and $H_0$ denotes the Hubble constant. 

Evidence for dark energy is found in the detection of an accelerated expansion of the Universe at high confidence levels 
in Type Ia supernova surveys \citep{rie98,per99,rie04,joh04}. An accelerated Hubble flow points to weak but cosmologically significant forces of non-baryonic origin with a zero-crossing redshift $z_t\simeq 0.4$ for the deceleration parameter $q=q(z)$, defined by $q=-H^{-2}\ddot{a}/a$ in a FRW universe with scale factor $a=a(t)$ and dynamical Hubble constant $H=\dot{a}/a$. Recent supernova surveys with enlarged number of supernova detections extending to redshifts of order unity resolve the Taylor series expansion $q(z)=q_0+(dq/dz)_0 z$ \citep{rie04}, that may serve as a constraint on theories of static or dynamic dark energy. 

An excess of cold dark matter to baryonic matter appears by a factor of about six cosmologically in the standard model of $\Lambda$ and cold dark matter ($\Lambda$CMD, \cite{bah99,ade13}), which appears pervasive across essentially all scales substantially greater than the solar system (e.g. \cite{tri14}). Notably, it appears in galaxy clusters (by a factor of about eight \citep{gio09}), in the \cite{fab76} and \cite{tul77} relations for stellar motion
in galaxies, in globular clusters \citep{her13} and in ultra-wide stellar binaries \citep{her12}.

There is a notorious tension between dark energy measurements in data of the cosmic microwave background (CMB)
and Type Ia supernovae, covering high and, respectively, low redshift epochs of the Universe. Recently, \cite{sal14} 
address the possibility of a dynamical dark energy. Based on CMB and supernova data, some tentative evidence 
is found favoring a possible interaction of dark energy with cold dark matter at relatively low redshift, while a constant 
dark energy in $\Lambda$CMD is ruled out at 99\% confidence level. 

In this Letter, we consider the remarkably universal appearance of dark energy as a consequence of the cosmological horizon as a Lorentz covariant boundary condition to general relativity. As a trapped surface, this apparent horizon defines the maximal size of holographic screens, whose surface area $A_H=4\pi R_H^2$, where $R_H$ denotes the Hubble radius. In setting the maximal radius of a holographic screen, it defines the number of degrees of freedom in the universe \citep{bek81,tho93,sus95}. The information represents the microphysical distribution of matter, which is presently nearly maximal within a factor of order unity \citep{van15a}. This encoding is commonly envisioned in bits stored in discrete Planck sized surface elements of area $l_p^2$, $l_p=\sqrt{G\hbar/c^3}$, where $\hbar$ denotes the reduced Planck constant. 

In a cosmological holographic interpretation \citep{tho93,sus95,eas11}, the aforementioned scale $L_0$ appears in 
a phantom pressure $p_0 = -{L_0}/{cA_H}$ in the universe enclosed by aforementioned 
cosmological holographic screen \citep{van15b}. By Lorentz invariance of the cosmological horizon, it is accompanied by a 
dark energy $\rho_\Lambda=-p_0$ (e.g., \cite{wei89}), manifest in an isotropically accelerated Hubble flow, whereby
\begin{eqnarray}
\Omega_\Lambda=\frac{2}{3},
\label{EQN_OML0}
\end{eqnarray}
where $\Omega_\Lambda = \rho_\Lambda/\rho_c$, $\rho_c = 3c^2H^2/8\pi G$. In what follows, we shall use geometrical units with
$G=c=1$ except when stated otherwise.

To study the observational consequences of (\ref{EQN_OML0}), we consider its implications for general relativity in an isotropic and homogeneous universe described by the FRW line-element. The specific back reaction to general relativity will be expressed in terms of modified FRW
equations. In obtaining a second order equation, it represents a singular perturbation of the original Hamiltonian energy constraint in
general relativity. It describes an evolution equation for the Hubble radius in response to the presence of the dimensionless matter 
density $\Omega_m$. 

We here study the implications for an evolving dark energy
in terms of the associated accelerated expansion and confronted with the transition redshift $z_t$, $q(z_t)=0$, 
and the observed confidence region of $(q_0,(dq/dz)_0)$ obtained from Type Ia supernova surveys. 
In the present approach, the dynamical dynamical dark energy is co-evolving with $\Omega_m$ with no direct interaction, different from
the proposed Ansatz in \cite{sal14}. We consider only the cosmology evolution after the radiation dominated epoch, since baryon 
nucleosynthesis is not affected in the present approach. 

\section{Holographic phantom energy}

In the approximation of a homogeneous and isotropic universe, we consider the FRW the line-element 3+1 $ds^2 = -dt^2 + a^2(t) h_{ij}dx^idx^j$ in spherical coordinates
\begin{eqnarray}
ds^2 = -dt^2 + a(t)^2\left( \frac{dr}{1-kr^2} + r^2 d\theta^2 + r^2\sin^2\theta\right).
\label{EQN_FRW}
\end{eqnarray}
Here, $r$ denotes the comoving radial coordinate scaled by $a(t)$ with dynamical Hubble parameter $H=\dot{a}/a$. Surfaces
of constant world-time $t$ have intrinsic Ricci scalar curvature $^{(3)}R=6k/a^2$, $k=\{-1,0,1\}$, and extrinsic curvature 
$K_{ij}=-a\dot{a}h_{ij}$. At a radius $R_H$, a congruence of outgoing
null-geodesics has vanishing expansion, where the unit space-like normal $s^i$ satisfies $D_is^i-K+s^is^jK_{ij}=0$ \citep{yor79,bau03}, i.e. $R_H={c}/{\sqrt{H^2+{k}/{a^2}}}$ (e.g. \cite{eas11}).
We shall consider a three-flat universe $(k=0$), as expected from primordial inflation \citep{lid00}.
The cosmological horizon assumes the radius $R_H=c/H$. 
We here consider a generalization of the de Sitter temperature \citep{gh77} given by the \cite{unr76} temperature of its surface gravity
(adapted from \cite{cai05})
\begin{eqnarray}
k_BT_{H}=\frac{H\hbar}{2\pi} \left( \frac{1-q}{2} \right),
\label{EQN_RH}
\end{eqnarray}
where $q=-a\ddot{a}/\dot{a}^2$ is the deceleration parameter and $k_B$ denotes the Boltzmann constant. 
As a null-surface, the cosmological horizon has a Bekenstein-Hawking entropy $S/k_B = (1/4)A_H/l_p^2$, here
identified with the number of degrees of freedom in the phase space in the visible Universe. 

In a holographic interpretation the de Sitter temperature (\ref{EQN_RH}) of the cosmological 
horizon introduces, by Lorentz invariance, a dark energy density from a phantom pressure \citep{eas11}
from virtual displacements in accord with Gibbs' principle (e.g. \cite{ver11,van12}),
\begin{eqnarray}
-p = A_H^{-1} T_{H} \frac{dS}{dR} = \frac{k_BT_{H}}{2R_Hl_p^2}.
\label{EQN_rL}
\end{eqnarray}  
With (\ref{EQN_RH}), (\ref{EQN_rL}) reduces to the equivalent local expression $p=-\ddot{A}/(2A)$
in terms of accelerated growth of the number of degrees of freedom in the comoving volume within
a comoving surface area $A(t,r)=4\pi a^2(t)r^2$ of constant $r$. 
With $q_0=-1$ of de Sitter space, (\ref{EQN_rL}) recovers (\ref{EQN_OML0}).
As a null-surface, the cosmological horizon is Lorentz invariant, whereby (\ref{EQN_rL}) introduces a dynamical dark energy
$\rho_\Lambda = -p$, satisfying  
\begin{eqnarray}
\Omega_\Lambda =\frac{2}{3}\left(\frac{1-q}{2}\right)
\label{EQN_OML}
\end{eqnarray}
as a generalization of (\ref{EQN_OML0}). 
For instance, $\Omega_\Lambda\simeq \{0,\frac{1}{2},\frac{2}{3},1\}$ in, respectively, 
a radiation dominated, matter dominated, present day and $\Lambda$ dominated epoch.

\section{Sixfold enhanced coupling to $\Omega_m$}

The Einstein equations for a universe with cosmological constant and stress-energy tensor $T_{ab}$ of matter are
\begin{eqnarray}
G_{ab} + \Lambda g_{ab} = 8\pi T_{ab},
\label{EQN_G}
\end{eqnarray}
where $G_{ab}$ denotes the Einstein tensor. Ignoring the holographic back reaction from the cosmological event horizon, we recall for a three-flat space the pair of FRW equations 
\begin{eqnarray}
\frac{\ddot{a}}{a} = \frac{1}{3}\Lambda -\frac{1}{2} H^2 \Omega_m 
\label{EQN_FRW1}
\end{eqnarray}
and
\begin{eqnarray}
\frac{\ddot{a}}{a} = -\frac{4\pi}{3}\left( \rho_m + 3 p\right) +\frac{8\pi}{3}\rho_\Lambda.
\label{EQN_FRW2}
\end{eqnarray}
Here $\Omega_m=\rho_m/\rho_c=\omega_m(a_0/a)^3$ is  
cold ($p=0$) dark matter of $\rho_m = \rho_0(a_0/a)^3$ and $\omega_m=\rho_0/\rho_c^0$ at present-day closure 
density $\rho_c^0=3H_0^2/8\pi$ defined by the Hubble constant $H_0$ at $z=0$; and $\Lambda=8\pi \rho_\Lambda$. Matched to CMB data, it  
gives a $\Lambda$CDM concordance with $\omega_m \simeq 0.31$ \citep{ade13}.
For $T_{ab}=\rho_m u_au_b + p(g_{ab}+u_au_b)$ describing matter with a total mass-energy density $\rho_m$ and unit velocity four-vector $u^b$ in the presence of a pressure $p$, (\ref{EQN_G}) implies more generally a Hamiltonian energy constraint $^{(3)}R-K:K+K^2=16\pi \rho_m + 16\pi \rho_\Lambda$ in a 3+1 line element with three-curvature $^{(3)}R$ and extrinsic curvature $K_{ij}$. For (\ref{EQN_FRW}) with $k=0$ and $K_{ij}=-a\dot{a}\delta_{ij}$, it implies \begin{eqnarray}
\Omega_m + \Omega_\Lambda=1,
\label{EQN_OM}
\end{eqnarray}
where $\Omega_m = \rho_m /\rho_c$. 

Our dynamical dark energy $\Lambda=8\pi\rho_\Lambda =H^2(1-q)$ in (\ref{EQN_G}), i.e.,
\begin{eqnarray}
G_{ab} = 8 \pi T_{ab} - H^2(1-q)g_{ab},
\label{EQN_MEE}
\end{eqnarray}
splits up into two parts: a source term $H^2$ and $-qH^2 = \ddot{a}/a$. Second order in time, the latter modifies the principle part of the Einstein tensor. Moving second-order terms on the left hand-side and leaving first-order terms on the right hand side, we have, for the FRW line-element under consideration, $\tilde{G}_{ab}=8 \pi T_{ab} - H^2g_{ab}$, where $\tilde{G}_{ab}=G_{ab} +(\ddot{a}/a) g_{ab}$. Specifically, (\ref{EQN_OML}) implies that the constraint (\ref{EQN_OM}) is now second-order in time, ${\ddot{a}}/{a} = 2H^2 - 8\pi \rho_m$, i.e.,
\begin{eqnarray}
\frac{\ddot{a}}{a} = 2H^2 - 3H^2\Omega_m.
\label{EQN_M1}
\end{eqnarray}
The general equation (\ref{EQN_M1}) is succinctly expressed in terms of an evolution equation for the
Hubble radius \footnote{Allowing for curvature, $\Omega_k = -k / aH^2$, the general expression is
$\dot{R}_H = 2\beta - 3 +3\beta^2\Omega_m$, where $\beta = HR_H/c=\sqrt{1-k(R_H/a)}$ $(k=\{\pm1,0\})$.} 
\begin{eqnarray}
\dot{R}_H = 1+q = -1 + 3\Omega_m,
\label{EQN_M2} 
\end{eqnarray}
A radiation dominated epoch has $q=1$ with $\Omega_m$ replaced by $\Omega_r = 1$, whereas a cold matter dominated epoch has $q=1/2$ with $\Omega_m = 5/6$ accompanied by dynamical dark energy fraction $\Omega_\Lambda=1/6$ with late time runaway solution $a={a_0}/{(t_*-t)}$
with $t^*$ on the order of a Hubble time. We have $q=-2$ and $\dot{R}_H=-1$ upon approaching $R_H(t_*)=0$. 

Remarkably, coupling to matter in (\ref{EQN_M1}) is {\em six times stronger} than in the original FRW equation (\ref{EQN_FRW1}). For 
purposes of numerical integration, we normalize it as 
\begin{eqnarray}
\frac{a^{\prime\prime}}{a} = 2 h^2 -3\omega_m\left(\frac{a_0}{a}\right)^3,
\label{EQN_M2B}
\end{eqnarray}    
where $a=a(\tau)$ as a function of dimensionless time $\tau=tH_0$ and $h=H/H_0$. 

On the right hand side of the second FRW equation (\ref{EQN_FRW2}), we have $ H^2\left[-\frac{1}{2} \Omega_m -\frac{3}{2} \Omega_p + \Omega_\Lambda\right] = H^2\left[-\frac{1}{2} - \frac{3}{2} \Omega_p + \frac{3}{2} \Omega_\Lambda\right]$, where $\Omega_ p = p/\rho_c$. By (\ref{EQN_OML}), it reduces to $8\pi p = -\ddot{a}/a$. Here, the appearance of a logarithmic acceleration $\ddot{a}/a$ appears natural in the face of the high symmetry of a FRW cosmology. In Gowdy $T^3$ cosmologies, for instance, polarized waves are described by a linear wave equation for the logarithm a diagonal metric \citep{ber93,van97}.
The result is a linear relation
\begin{eqnarray}
q=3\Omega_p,
\label{EQN_pq}
\end{eqnarray}
between deceleration and pressure. Since $q$ drops below zero at present, $p$ is currently negative. 
In our interpretation, $p$ has a corresponding positive two-dimensional pressure in cosmological holographic screens, whereby (\ref{EQN_pq}) defines a positive correlation between acceleration and pressure. It suggests that (\ref{EQN_pq}) is an inertial equation in cosmological evolution. While $p$ and dark matter are both evolving in time, (\ref{EQN_pq}) has no direct coupling between them. It appears that the total phantom pressure in the universe consists of a thermal component, due to (\ref{EQN_OML0}) and its extension (\ref{EQN_OML}), and the inertial component (\ref{EQN_pq}) with, at present, no small parameters.

\section{Accelerated expansion}

The modified FRW equation (\ref{EQN_M2}) has observational consequences for weak gravitational interactions on the scale $a_H$, that are in dramatic contrast to what is naively expected based on (\ref{EQN_FRW1}). Numerical integration of (\ref{EQN_M2B}) gives
a graph $q_0(z)$, independent of $H_0^2\omega_m$, since different choices for $H_0^2\omega_m$ in (\ref{EQN_M2}) can be absorbed in a rescaling of time. Fig. 1 shows a transition redshift $q(z_t; q_0)=0$ as a function of the deceleration parameter at the present epoch $z=0$, satisfying
\begin{eqnarray}
z_t (q_0) = 0.43 - 0.24(1+q_0).
\label{EQN_zt}
\end{eqnarray}
This is a model independent result consistent with the observed values $z_t = 0.46\pm0.13$ and $q_0(0)\simeq -0.8$ inferred from 
the gold and silver sample in the supernova survey of \cite{rie04}.
\begin{figure} 
\centerline{\includegraphics[width=90mm,height=80mm]{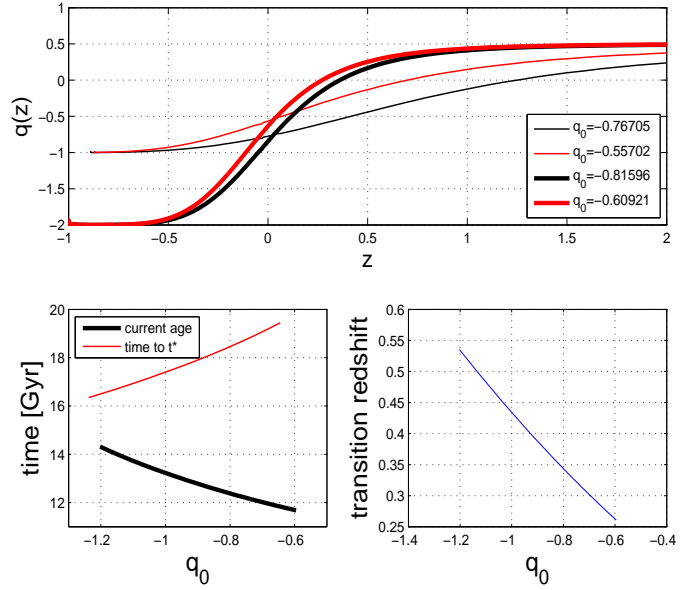}}
\caption{Evolution of the deceleration parameter $q(z)$ of a three-flat cosmology as a function of redshift $z$ (top) according to
general relativity with (thick lines) and without (thin lines) back reaction from the cosmological horizon. As function of $q_0=q(0)$, the age of the universe, the time remaining to $t^*$ (left bottom) and the transition value $z_t$ at crossing $q(z_t; q_0)=0$ (right bottom).}
\label{fig1}
\end{figure}

To facilitate our confrontation of with data in the $(q,(dq/dz)_0)$ plane, we consider $(q,(D_h^+q)_{z_0})$ using the approximation one-sided finite difference
\begin{eqnarray}
D_h^+ q(z_0) = \frac{q(z_0+ h )-q(z_0)}{\Delta z} = (dq/dz)_{z_0} + O\left(h \right)
\label{EQN_Dh}
\end{eqnarray}
for $z_0$ close to zero and moderate $h=\Delta z$, permitted by the redshift range of the data. Even though measurement of $q(z)$ is quite challenging, taking $0<\Delta z<1$ is expected to allow for a reasonable estimate with moderate dependence on choice of $\Delta z$ for
both model alternatives. Here, (\ref{EQN_Dh}) is calculated using a linear fit by the method of least square error. Results for different
values of $h$ serve to indicate the window of uncertainty in the extraction of $dq/dz$ from the data.

\begin{figure} 
\centerline{\includegraphics[width=85mm,height=80mm]{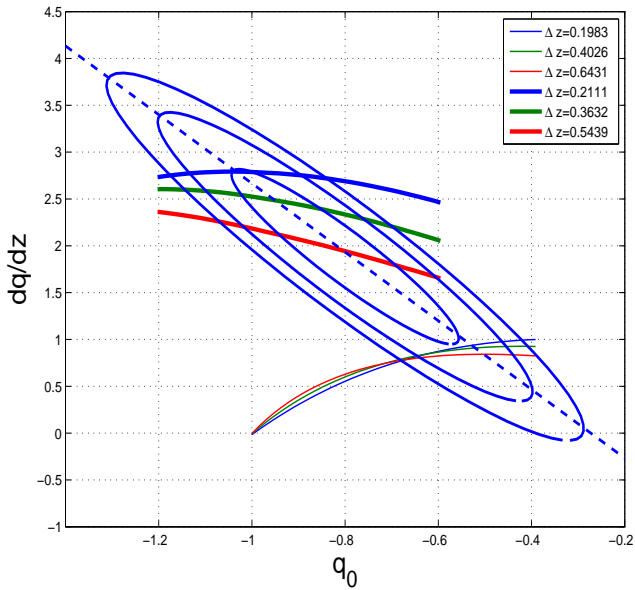}}
\caption{Model graphs $(q_0,D_h^+q(z_0))$ $(z_0=0.008)$ are compared with confidence regions 
(1, 2 and 3$\sigma$ ellipses) of the Type Ia supernovae survey (gold and silver sample) 
of Riess et al. (2004) for a three-flat cosmological evolution according to 
general relativity with (thick lines) and without (thin lines: $\Lambda$CDM) back reaction from the cosmological event horizon. 
Shown are the one-sided finite-difference estimates $D_h^+q(z_0)$ for $dq/dz$ for various choices of $h=\Delta z$,
in simulated extraction from data from nearby events. The curves shown for different choices of $h$ point to a window 
of about $-1<q_0<-0.7$.}
\label{fig1b}
\end{figure}

Fig. 2 shows our observational test of (\ref{EQN_M1}) and (\ref{EQN_FRW1}) with evolving, respectively,
static dark energy in $\Lambda$CMD for a three-flat cosmology \citep{ade13}, overlaid with the observed 
confidence region in the $(q_0,(dq/dz)_0)$ plane obtained from the supernova survey of \cite{rie04}. 
The latter points to a deceleration parameter value about $q_0\simeq -0.8$ in the range $-1 < q_0 <-0.7$ of \cite{joh04}
(see also \cite{tri06}). In this particular range of $q_0$, Fig. 2 shows a satisfactory agreement between data and (\ref{EQN_M1}),
while ruling out (\ref{EQN_FRW1}) at 3 $\sigma$.

\section{Conclusions}

A holographic back reaction of the cosmological horizon is shown to give a singular perturbation of the FRW equations, described by 
a dynamical dark energy and a sixfold enhancement in coupling to matter. The observational consequences present a striking departure to 
what is expected based on general relativity alone. Without fine-tuning, these consequences are in agreement with observations, on the lifetime
of the Universe and the transition redshift at zero-crossing of the deceleration parameter (Fig. 1) as well as confidence intervals on the
latter and its first derivative (Fig. 2). 

While Fig. 2 shows a tension with $\Lambda$CDM in the full sample of gold and silver, this appears less so
in gold alone. It seems prudent to improve our understanding of these samples in a future analysis.

We emphasize that in a preceding radiation dominated era, $q=1$, $\Omega_\Lambda=0$ in (\ref{EQN_OML}) and there is no 
radiative correction to the Einstein tensor. As nucleosynthesis takes place in a radiation dominated era,
its results are left unchanged in the present theory. 

As holographic quantities, our dark energy and dark matter appear uniformly on a cosmological scale in (\ref{EQN_M2}), 
driving the overall evolution of the universe in co-evolution with but without direct interaction with baryonic matter. 
Of holographic origin, $\Lambda$ in (\ref{EQN_MEE}) represents a Lorentz invariant contribution from the
dynamical evolution of the number of degrees of freedom in comoving phase space. Based on these theoretical
and observational consequences, holographic cosmology gives
some novel justification for a holographic origin of the observed
three-dimensional phase space (Bekenstein 1981; Õt Hooft 1993;
Susskind 1995).

\mbox{}\\
{\bf ACKNOWLEDGMENTS}\\ 
\mbox{}\\
The author gratefully acknowledges the anonymous referee for constructive comments which
have considerably improved the readability of this letter and 
G. 't Hooft and X.-D. Li for stimulating discussions.

\label{lastpage}

\end{document}